\documentclass[11pt]{article}
\usepackage{moriond,epsfig}

\def\Journal#1#2#3#4{{#1} {\bf #2}, #3 (#4)}

\def\NIM{\em Nucl. Instrum. Methods}

\def\PLB{{\em Phys. Lett.}  B}
\def\PRL{\em Phys. Rev. Lett.}
\def\PRD{{\em Phys. Rev.} D}

\def\be{\begin{equation}}
\def\ee{\end{equation}}
\def\bea{\begin{eqnarray}}
\def\eea{\end{eqnarray}}

\begin{document}
\vspace*{4cm}
\title{TOP PRODUCTION PROPERTIES AT THE TEVATRON}

\author{ ANTHONY VAICIULIS }

\address{For the CDF and D0 Collaborations\\
Department of Physics and Astronomy, University of Rochester,\\ 
Rochester, NY 14627 USA}

\maketitle\abstracts{
Following the confirmation of the top quark discovery at the Tevatron, the
next step is to begin studying its properties. Because the scale of electroweak
symmetry breaking is of the same order as the measured top mass, the top
quark may be very sensitive to new physics. We describe measurements of
the total $t\bar{t}$ cross section, the $t\bar{t}$ invariant mass, the
top transverse momentum distribution, and single top production from 
Run 1 data at D0 and CDF. All of these measurements will become more
precise in Run 2.}

\section{Introduction}
The top quark is interesting for many reasons, not the least of which is that 
it is about forty times more massive than the next heaviest quark, the b quark.
One of the great problems of particle physics today is to understand
electroweak symmetry breaking and the role the top quark plays in this.
Now that the direct observation of the top quark has been confirmed, it 
is time to study its properties to help answer this question.

All direct measurements of the top quark are from the Fermilab $p\bar{p}$ 
collider Run 1.\cite{cdfd0findtop} The Tevatron delivered collisions
at a center-of-mass energy of 1800 GeV and an integrated luminosity of about 
0.1 fb$^{-1}$,
between 1993 and 1996. Accelerator and detector upgrades are in progress for 
Run 2, which will have a center-of-mass energy of
2000 GeV resulting in a 40\% increase in $\sigma(t\bar{t})$. It will have an 
integrated
luminosity of about 2 fb$^{-1}$, about twenty times larger than the Run 1 
data sample.
The effective increase in the size of the data sample will be even higher because
of greater b tagging efficiency due to new silicon vertex detectors at CDF 
and D0.

According to the Standard Model,
the top quark is produced at the Tevatron mainly via $t\bar{t}$ pair production. 
The two main
processes are the $q\bar{q}$ annihilation diagram ($q\bar{q} \rightarrow t\bar{t}$) and gluon-gluon
fusion ($gg \rightarrow t\bar{t}$). These contribute about 90\% and 10\% 
respectively to the
$t\bar{t}$ cross section at the Tevatron, which has a Standard Model predicted
value of 4.7-5.5 pb. \cite{xsectiontheory}
The expected $t\bar{t}$ cross section of $\sim$5.0 pb is a tiny fraction
of the total cross section; from
more than 10$^{12}$ $p\bar{p}$ collisions in Run 1, the top measurements are based
on $\sim$100 events.

The measurements of top have been made with D0 and CDF, \cite{cdfd0detector} 
general purpose detectors designed to study high-P$_T$ interactions. 
D0 has excellent calorimeter resolution and
good muon coverage while CDF has emphasized tracking, with a magnetic field and
a silicon vertex detector (SVX).
There are many improvements to both detectors for Run 2. For D0 these include
acquiring a central magetic field and an SVX. CDF will have an improved
silicon detector system, able to do stand-alone tracking.

In order to identify events in which top quarks are produced, the decay
products must be detected. According to the Standard
Model, the top quark decays to Wb with a branching ratio of nearly 100\%.
The individual branching
ratios are 
BR(W$^{+} \rightarrow e^{+}\nu$) = 1/9,
BR(W$^{+} \rightarrow \mu^{+}\nu$) = 1/9,
BR(W$^{+} \rightarrow \tau^{+}\nu$) = 1/9,
BR(W$^{+} \rightarrow q\bar{q}$) = 6/9.
This leads to four main $t\bar{t}$ event topologies:
dilepton (5\%), lepton +jet (30\%), all-hadronic (44\%) and events with taus 
(21\%).
The dilepton channel is the most pure but has the fewest number of events because
of the low branching ratio. The all-hadronic channel has large QCD backgrounds
($p\bar{p} \rightarrow$ six jets).

\begin{figure}
\psfig{figure=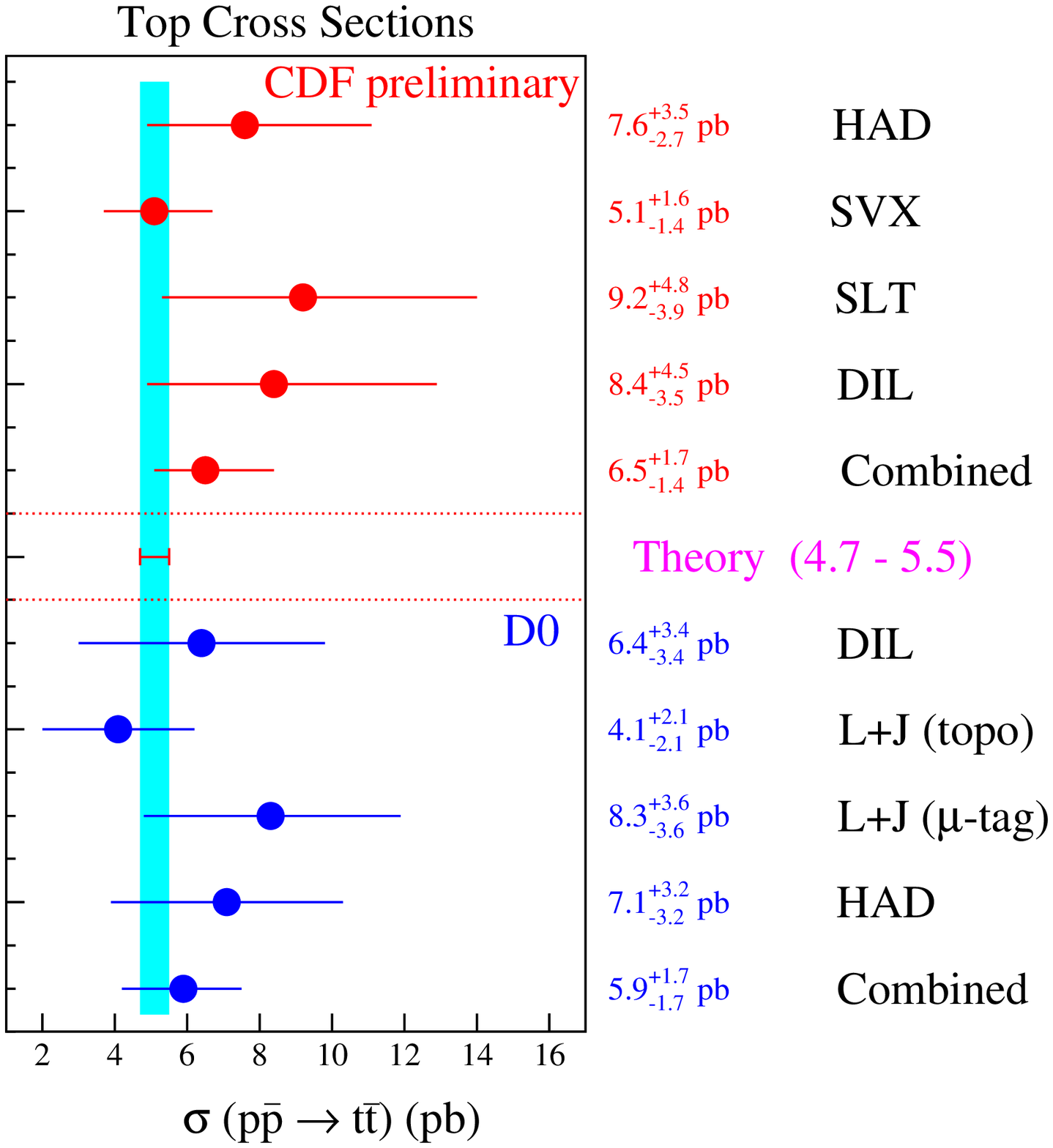,height=3.0in}
\hspace*{1cm}
\psfig{figure=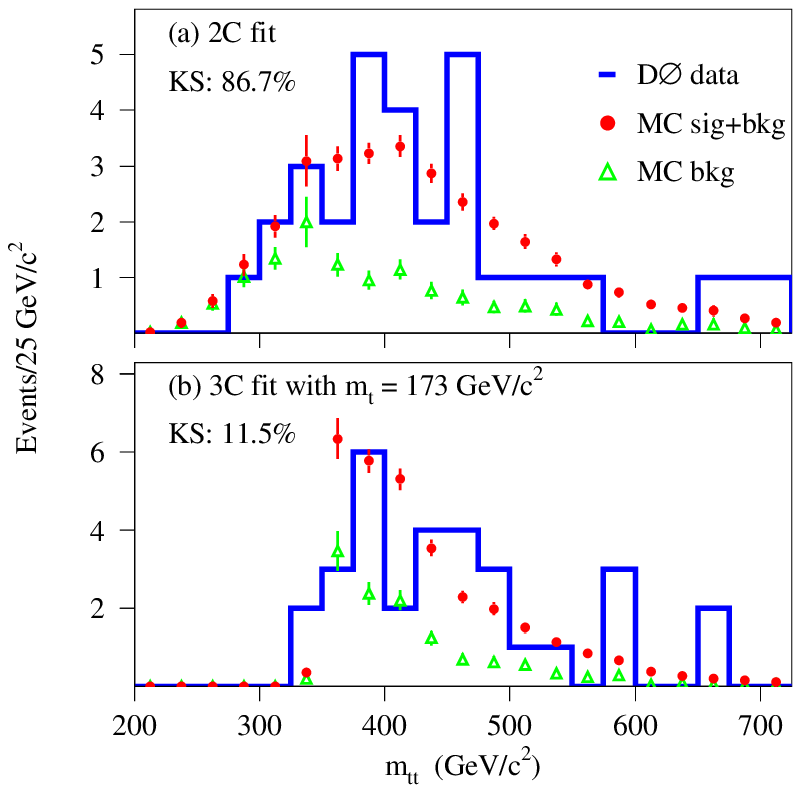,height=3.0in}
\caption{(a) Measured $t\bar{t}$ cross section for D0 and CDF in the various 
channels compared to the theoretical prediction (band). (b) The observed 
M$_{t\bar{t}}$ spectrum (histogram) from D0 using a 2-contraint and 3-constraint 
fit and compared to the expected background + $t\bar{t}$ signal.
\label{fig:xsection}}
\end{figure}

To reduce backgrounds, two different methods are used to tag
jets originating from b quarks. Both D0 and CDF use the `soft lepton tag'
which seeks to identify
the soft lepton in semileptonic b or c decays ($b \rightarrow cl\nu$ or 
$b \rightarrow c \rightarrow sl\nu$). CDF also attempts to tag b jets by
finding secondary vertices using the SVX.
The efficiencies of the `soft lepton tag' and SVX methods are $\sim$20\% and
$\sim$50\% respectively.

\section{Total $t\bar{t}$ Cross Section}
A first indication of new physics might be seen when measuring 
$\sigma(t\bar{t})$ and comparing it to the expectation from the Standard
Model. A disagreement could indicate non-SM physics including
a heavy resonance decay into a $t\bar{t}$ pair, a non-SM decay of top (into
supersymmetric particles, for example) or an unexpected branching ratio
for a particular top decay channel. The CDF and D0 cross section measurements
for the different channels are shown in figure~\ref{fig:xsection}(a). 
The CDF combined cross
section measurement of $6.5^{+1.7}_{-1.4}$ pb \cite{cdfxsection} 
agrees with the D0 measurement of
5.9 $\pm$1.7 pb. \cite{d0xsection} Both measurements from Run 1 data agree within 
uncertainties with the Standard Model prediction (shown as a band) of about 
5.0 pb.

\section{Invariant Mass of $t\bar{t}$ and top P$_T$}
\begin{figure}
\psfig{figure=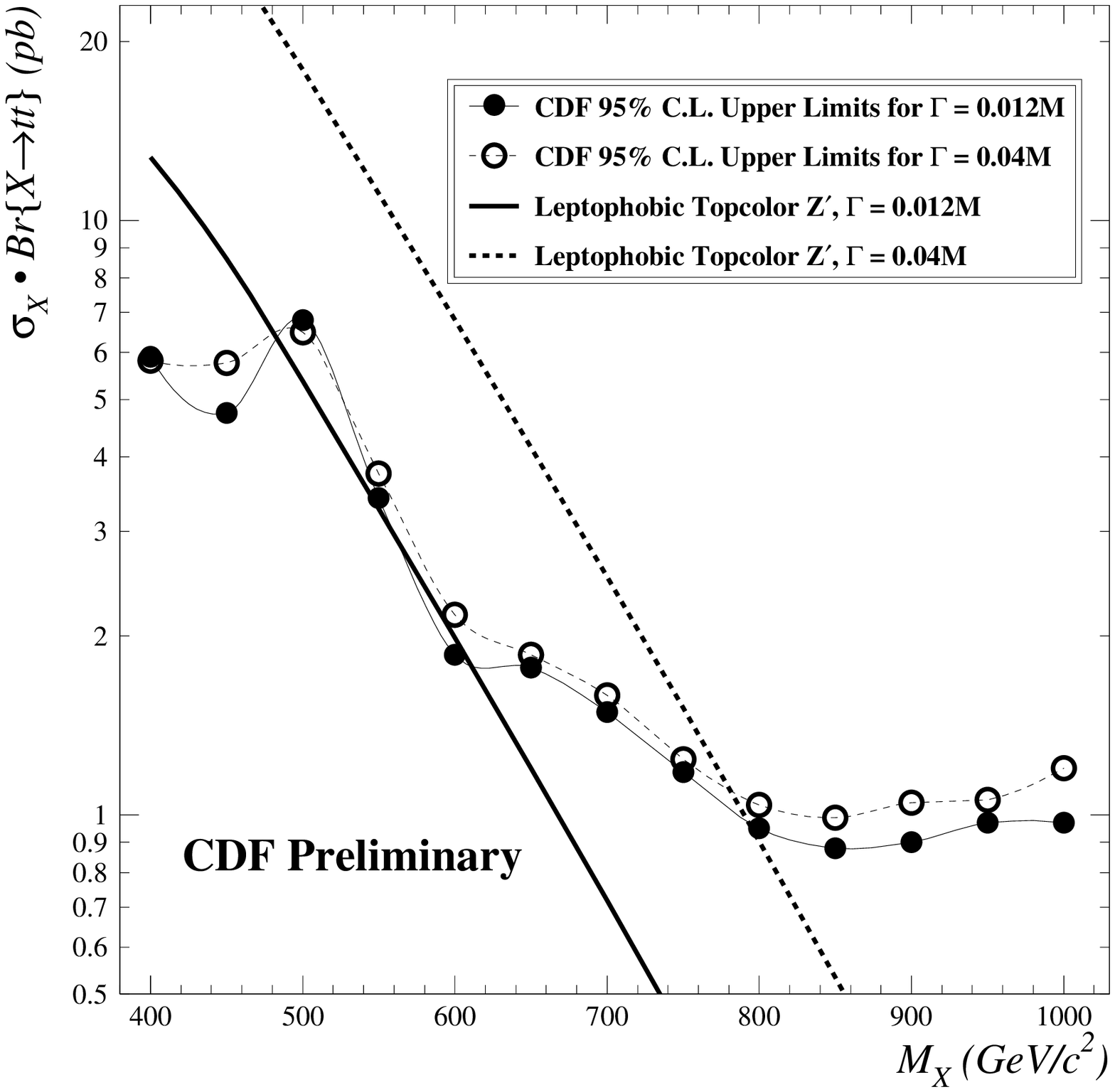,height=2.3in}
\hspace*{0.1cm}
\psfig{figure=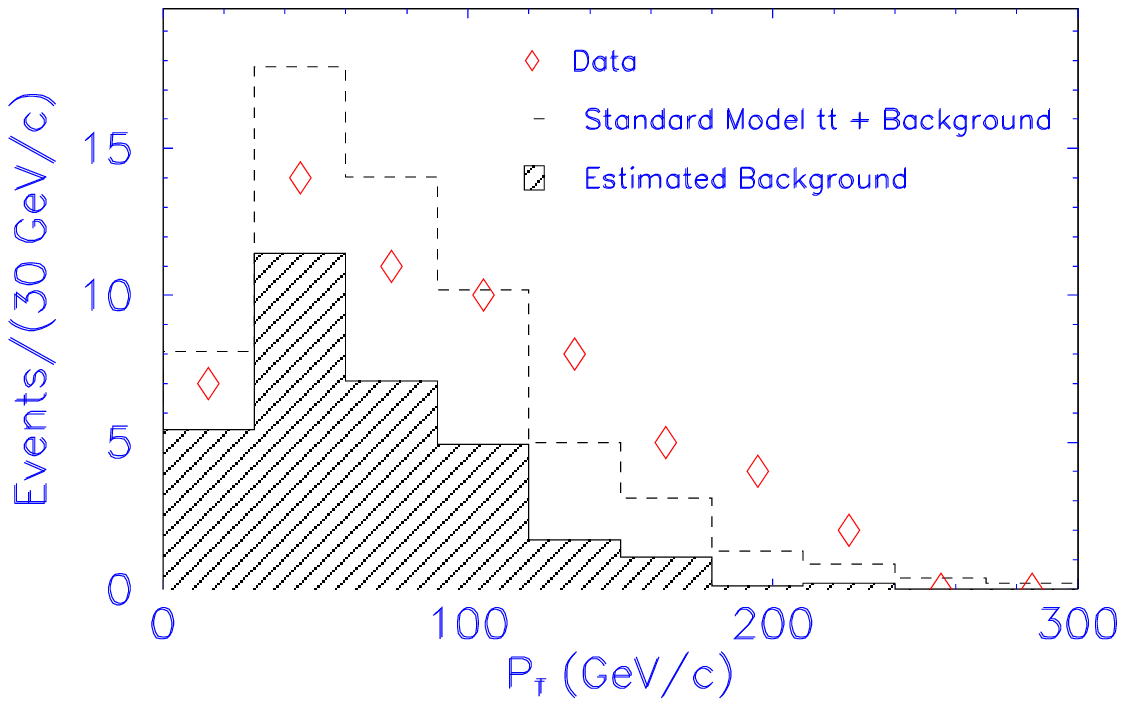,height=2.4in}
\caption{(a) The 95\% CL upper limits on 
$\sigma(p\bar{p} \rightarrow X)\times$BR($X \rightarrow t\bar{t})$ 
as a function of mass (solid and open points) compared to the cross section
for a leptophobic topcolor Z' for two resonance widths (CDF).
(b) The measured P$_T$ distribution for hadronically-decaying top quarks
compared to the Standard Model expectation (CDF).
\label{fig:mtt_limits}}
\end{figure}

A study of the $t\bar{t}$ invariant mass spectrum is one way to probe for new
physics. The spectrum may disagree with that predicted by the Standard Model if
a $p\bar{p}$ produced heavy object decays into a $t\bar{t}$ pair. 
The topcolor-assisted technicolor model,
\cite{topcolor} for example, predicts resonances in the M$_{t\bar{t}}$ spectrum: a
narrow topcolor Z' and a wide top gluon.
Figure \ref{fig:xsection}(b) shows the $t\bar{t}$ invariant mass spectrum as 
measured by D0. The data are consistent with the expected $t\bar{t}$ signal 
plus background. 
CDF measured the $t\bar{t}$ invariant mass spectrum and placed upper limits on 
$\sigma(p\bar{p} \rightarrow X)\times$ BR($X \rightarrow t\bar{t})$.
\cite{cdfmttbar}
The M$_{t\bar{t}}$
data is fit to templates of $t\bar{t}$, W + jets (to model background) 
and $Z' \rightarrow t\bar{t}$ (an example narrow resonance). 
The 95\% CL limit is shown
versus the mass of the heavy object, M$_X$, in figure ~\ref{fig:mtt_limits}(a). 
A topcolor Z'
with a width of 0.04M$_{Z'}$ (0.012M$_{Z'}$) is excluded for masses below 780 GeV 
(480 GeV).
In Run 2 it is hoped to be able to place limits on such narrow resonances
up to $\sim$1 TeV. 

A model independent way of probing for new physics while studying top is
to measure the top quark transverse momentum distribution. In some theories
of non-SM phenomena this distribution can be significantly modified. CDF
uses a likelihood method to correct for reconstruction and resolution effects
to transform the observed top P$_T$ to the `true' top P$_T$. \cite{cdftoppt} 
Figure ~\ref{fig:mtt_limits}(b) shows the resulting distribution.
The data points are compared to the Standard Model $t\bar{t}$ plus background.
In this analysis, the data is divided into four bins of true P$_T$. The fraction of
events observed in each bin agrees within errors with the SM prediction.
Some models for non-standard production predict an enhancement of the 
top cross section for high-P$_T$ top quarks. This analysis has set
a 95\% CL upper limit of 0.16 for the fraction of events with P$_T > 225$ GeV, 
which seems to disfavor models predicting a distribution greatly enhanced 
at high-P$_T$.

\section{Single Top Production}
In addition to $t\bar{t}$ pair production, the Standard Model also predicts 
top to be produced at the Tevatron by means of single top production. 
The two main contributing processes
are Wg fusion ($\sigma = 1.7 \pm 0.3 $pb) and W* production 
($\sigma = 0.7 \pm 0.1 $pb ). Whereas the study of
$t\bar{t}$ pair production tells us about the strong interaction of the top quark,
single top production can help us learn about the electroweak interactions
of the top quark because of the Wtb coupling. The single top cross section
is proportional to $\Gamma(t \rightarrow Wb)$ which is proportional to 
the square of
the V$_{tb}$ CKM matrix element. Thus a measurement of $\sigma$(single top) can
be used to determine $|$V$_{tb}|$ without assuming a unitary three generation
CKM matrix as other measurements have done. If the $|$V$_{tb}|$ measurement 
is very different from unity, it could indicate non-SM physics such as a 
fourth generation.

A measurement of the single top cross section is difficult because of
the small cross section and the large W + two jets QCD backgrounds 
($Wb\bar{b}, Wc\bar{c}$).
A CDF analysis has used the lepton + jets channel.\cite{singletop}
It requires M$_{l\nu b}$ to be near
M$_{top}$ and fits the H$_t$ distribution (E$_t$ sum over jets, 
lepton, and missing E$_T$ in event)
to the expected QCD background, $t\bar{t}$ background, and single top signal. 
The expected background is much larger than the expected
signal, so the cross section is not measured from Run 1 data. Rather, an upper
limit on the cross section is determined. The preliminary CDF (D0) 95\% CL 
upper limit (both
processes combined) is 13.5 pb (47 pb). With the 
acquisition of an SVX for Run 2, the D0 measurement should be 
competitive with CDF. Run 1 limits
are much higher than the standard model prediction of $\sim$2.4 pb. Neural net
analyses are in progress in an attempt to handle the large backgrounds.
In Run 2 it is hoped to measure the single top cross section to a precision
of about (20-30)\%.

\section{Conclusions}

The discovery of the top quark in Run 1 at the Fermilab Tevatron was
a great success.
The studies on the small number of top events available in Run 1
data included measurements of the $t\bar{t}$ cross section, 
the $t\bar{t}$ invariant 
mass spectrum, the top quark P$_T$ distribution, and single top production. 
There has been no evidence for non-Standard
Model physics. Because of a factor twenty increase in luminosity and upgraded
detectors, there will be a greater potential to discover new physics
in Run 2 with higher precision top quark studies.

\section*{Acknowledgments}
This work was made possible by the support of the members of the CDF
and D0 collaborations and by the U.S. Department of Energy 
(grant DE-FG02-91ER40685).

\end{document}